\documentclass[twocolumn,showpacs,preprintnumbers,amsmath,amssymb]{revtex4}
\usepackage{graphicx}
\usepackage{dcolumn}
\usepackage{bm}
\usepackage{amssymb}
\usepackage{amsmath}
\begin{document}
\title{Penetration depth of electron-doped infinite-layer Sr$_{0.88}$La$_{0.12}$CuO$_{2+x}$ thin films}

\author{L. Fruchter, V. Jovanovic$^*$, H. Raffy, Sid Labdi$^+$, F. Bouquet and Z.Z. Li}%
\affiliation{Laboratoire de Physique des Solides, Univ. Paris-Sud, CNRS, UMR 8502, F-91405 Orsay Cedex, France}
\affiliation{$^*$Institute of Physics, Pregrevica 118, Belgrade, Serbia}
\affiliation{$^+$Laboratoire des milieux nanom\'{e}triques, Universit\'{e} d'Evry, 91025 Evry cedex, France}
\date{Received: date / Revised version: date}

\begin{abstract}{The in-plane penetration depth of Sr$_{0.88}$La$_{0.12}$CuO$_{2+x}$ thin films at various doping obtained from oxygen reduction has been measured, using AC susceptibility measurements. For the higher doping samples, the superfluid density deviates strongly from  the $s$-wave behavior, suggesting, in analogy with other electron-doped cuprates, a contribution from a nodal hole pocket, or a small gap on the Fermi surface such as an anisotropic $s$-wave order parameter. The low value of the superfluid densities, likely due to a strong doping-induced disorder, places the superconducting transition of our samples in the phase-fluctuation regime.}
\end{abstract}


\maketitle

The question whether superconductivity obtained by doping the CuO$_2$ planes in hole-doped and electron-doped cuprates involves the same mechanisms is still a matter of debate. Indeed, evidences for asymmetry of the electronic properties between electron- and hole-doped compounds have  been pointed out long ago, some of them still controversial. 

First, the antiferromagnetic (AFM) order common to both systems at very low doping has often been reported to extend to a  much higher doping range  for electron-doped materials and found to overlap with the superconducting dome\cite{Dai2005}. However, this description is challenged by recent neutron-diffraction studies that conclude that genuine long-range antiferromagnetism and superconductivity do not coexist\cite{Motoyama2007}. On a theoretical point of view, a phase separation into a mixed antiferromagnetic and  superconducting (SC)  phase has been predicted for both classes of materials (although with a much larger energy scale in the case of hole-doped)\cite{Aichhorn2006}, while several experimental findings could be interpreted within a model that assumes coexisting AFM and SC orders\cite{Das2007,Das2008}. 

Then, one of the essential characteristics of the hole-doped cuprate superconductivity is the $d$-wave symmetry of its  order parameter, believed to reflect the pairing mechanism. In the electron-doped case (e-doped case), $d$-wave symmetry has been evidenced  by several high-quality experimental contributions,  however several others point towards a dominant $s$-wave order parameter (for a review, see Ref.~\onlinecite{Khasanov2008}). Recently, it has been  proposed that  this complexity may originate from the fact that, although e-doped cuprates properties for samples below optimal doping are indeed dominated   by the electron pockets of the Fermi surface, hole pockets are  developing  as the doping is increased and  may actually become dominant. The interplay between the doping evolution of the Fermi surface and a $d$-wave order parameter would then yield the rich behavior as a function of doping of the e-doped family\cite{Das2007,Das2008}. Some authors go further and suggest that, in the case of e-doped Pr$_{2-x}$Ce$_x$CuO$_4$ (PCCO), electrons may have no role in the occurrence of superconductivity, which would then be entirely dominated by the contribution of the hole pockets \cite{Dagan2007}. 

Confronted to this debated situation, experimental clues brought by an additional member of the restricted e-doped family -- the so-called `infinite phase' Sr$_{1-x}$La$_x$CuO$_2$ (SLCO) -- may prove useful. Concerning the issue of the  order-parameter symmetry, there have been several experimental investigations  for this material, most of them pointing towards a dominant $s$-wave superconducting order in the case of optimally-doped SLCO: the lack of a momentum dependence, as well as of a zero bias conductance peak, in tunneling spectroscopy\cite{Chen2002}; the temperature and magnetic field scaling of the mixed state specific heat\cite{Liu2005}; the local field distribution from low-angle neutron diffraction by the flux-line lattice\cite{White2008}; the muon-spin-rotation measurements of the flux-line-lattice field distribution\cite{Khasanov2008}. However,  other measurements found a temperature or a magnetic field dependence indicative of nodes in the gap\cite{Shengelaya2005,Satoh2008}. It was also pointed out that, in a similar way to what is observed for other e-doped cuprates, the zero-temperature superfluid density in SLCO does not follow the `universal' Uemura line for optimally hole-doped cuprates\cite{Uemura1991,Shengelaya2005,Satoh2008}, due to much shorter penetration depth for  comparable superconducting-transition temperatures. This tends to indicate that, for the e-doped cuprates, the Fermi liquid regime extends over a larger doping range than for the  hole-doped  cuprates\cite{Satoh2008}, in apparent contradiction with antiferromagnetism extending further into the underdoped regime.

\begin{figure}
\resizebox{0.9\columnwidth}{!}{\includegraphics{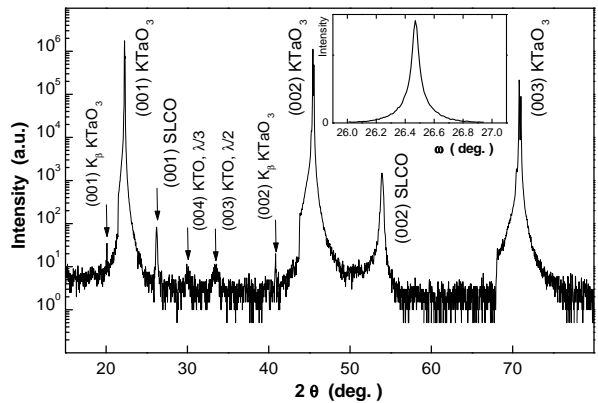}}
\caption {Cu-K$\alpha$ X-ray $\theta$--$2\theta$ diffraction scan, for a 590~\AA{} thick film. The inset is the rocking curve for the (002) peak, showing a mid-height width of 0.07 degree.}\label{lambda9}
\end{figure}

In the present study, we report measurements of the magnetic penetration depth of SLCO thin films. Despite its structural simplicity -- CuO$_2$ layers alternating with Sr$_{1-x}$La$_x$ layers -- SLCO is difficult to fabricate. As a bulk material, it can only be synthetized under pressure, and no single crystal could be grown up to now. As a thin film, epitaxial growth of $c$-axis oriented SLCO was however made possible by the use of the appropriate substrate\cite{Karimoto2004,Li2009,Jovanovic2009}. We have grown by rf magnetron sputtering several Sr$_{1-x}$La$_x$CuO$_2$ ($x=0.12$) thin films, approximately 400~\AA\ thick on $5\times 5$~mm$^2$ (100) KTaO$_3$ substrates. 
The CuO$_2$ planes doping with electrons is provided here both by the Sr$^{2+}$/La$^{3+}$ substitution (which was kept constant in this study), but also by an oxygen content reduction; indeed, during the process additional oxygen atoms enter the structure, most probably within the Sr$^{2+}$/La$^{3+}$ planes. The samples were annealed in-situ during the cooling procedure after deposition; the final doping state of a sample is determined by the temperature of annealing or its conditions (under vacuum or Argon pressure), the lower oxygen content resulting in a higher doping.
In a last step, a cover -- approximately 100~\AA~ thick -- of amorphous, insulating, material was deposited, to insure optimum stability of the film. We obtained thin films with $T_c$ up to 19~K for the lower oxygen content. Using  Cu-K$\alpha$ X-ray diffraction, the thickness of each film was measured from the low-angle Kiessig fringes, and the absence of parasitic phases was checked from conventional $\theta$--$2\theta$ pattern (see figure~\ref{lambda9}).

\begin{figure}
\resizebox{0.9\columnwidth}{!}{\includegraphics{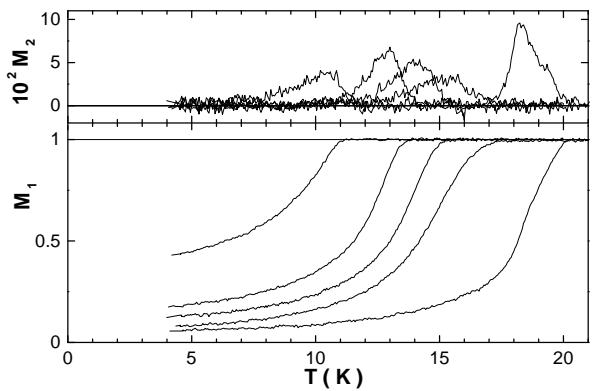}}
\caption {Normalized in-phase ($M_2$) and out-of-phase ($M_1$) pickup signal for films with $T_c$ = 19 K, 15 K, 14 K, 13 K, 11 K.}\label{lambda5}
\end{figure}

The penetration depth  $\lambda$  was measured using an ac susceptometer setup based on  Ref.~\onlinecite{Fiory1988}. It was built using two identical astatically wound pairs, each made from 1.25 mm  diameter, 100 turn coils. The use of quadrupoles minimizes the sample finite-size  contribution to the background signal, which was measured using a thick Nb sample with  the same dimensions as the measured films. The mutual inductance ($M_1+i\,M_2$) corrected from the finite-size effects was obtained using the procedure described in Ref.~\onlinecite{Turneaure1998} (Fig.~\ref{lambda5}). The geometry of the coils, combined with the relatively small thickness of the measured films (thinner than $\lambda$), allowed for an accurate determination of the mutual inductance on the whole temperature range. This  is necessary for a reliable determination of the penetration-depth temperature dependence: at the lowest temperature that could be reached by our apparatus (4~K), the out-of-phase signal was always larger than about 6\% of the signal above $T_c$, and twice as large as the correction brought by the background signal. The latter was independent of the temperature in our range:  the temperature of the measurement setup was kept  constant and independent of the sample temperature during measurement. The value of the  ac field at 50~kHz was adjusted in order to remain in the linear regime, where the measured inductance is independent of the excitation of the driving coil. Using a lookup table computed for the specific geometry of our setup, the penetration depth was obtained from the complex mutual inductance.  $T_c$ was obtained from a linear extrapolation of $\lambda^{-2}(T)$ to zero. 

\begin{figure}
\resizebox{0.9\columnwidth}{!}{\includegraphics{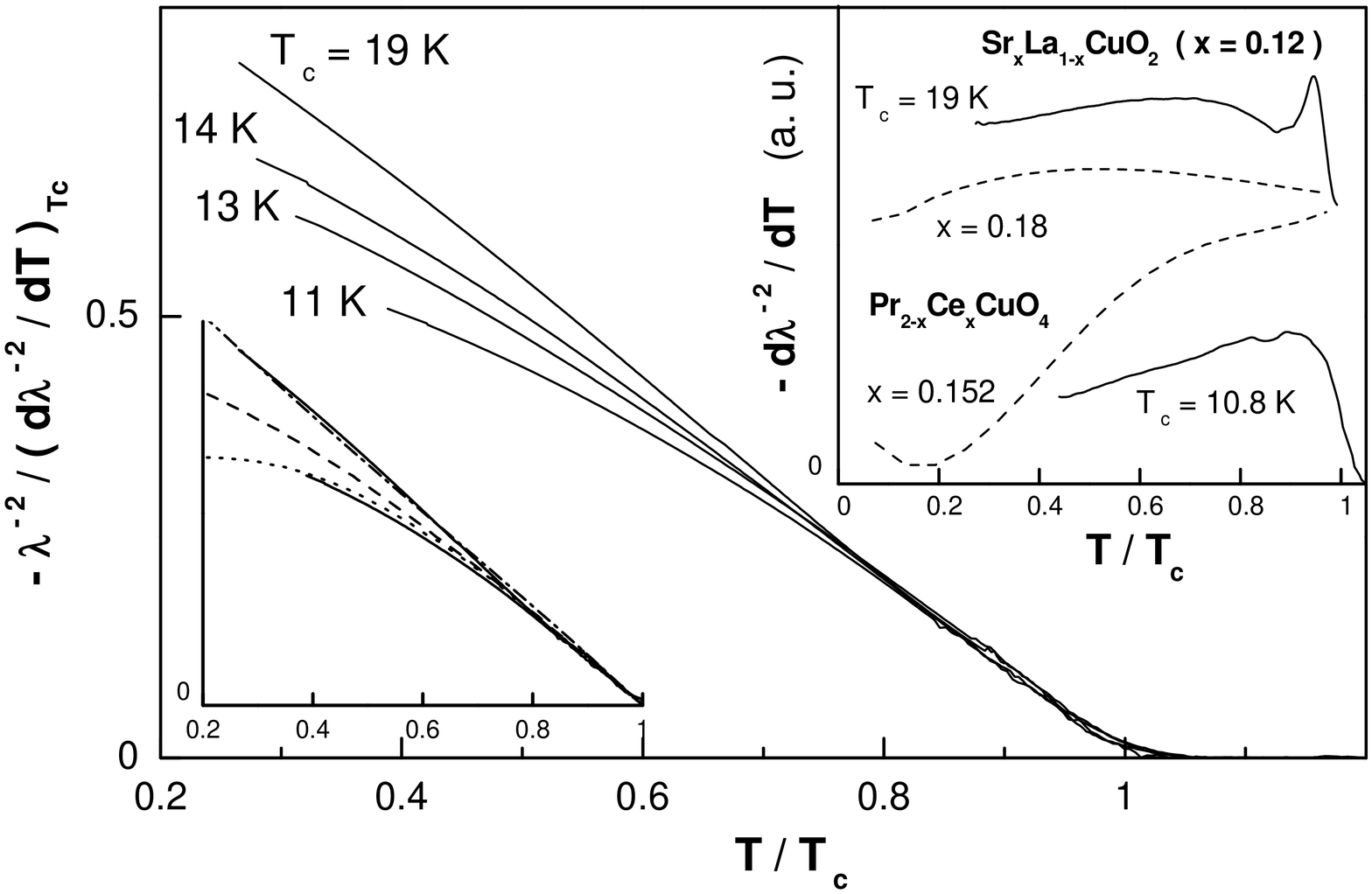}}
\caption{Squared inverse of the penetration depth, as obtained from the mutual inductance in Fig.~\ref{lambda5}, normalized to  the slope at $T_c$. The bottom inset is an attempt to fit the $T_c$ = 19 K data (upper full line) and the $T_c$ = 11 K data (lower full line) to clean $s$-wave (dotted line), $d$-wave (dashed line) and anisotropic $s$-wave (dashed dotted) theory. The top inset shows the temperature derivative for these two same samples (full lines); for comparison data from Ref.~\onlinecite{Das2008} are also plotted, corresponding to an optimally doped ($x = 0.152$) and overdoped ($x = 0.18$) Pr$_{2-x}$Ce$_x$CuO$_4$.}\label{lambda2}
\end{figure}

As can be seen in Fig.~\ref{lambda2}, the curvature for $\lambda^{-2}(T)$ is found to decrease as doping increases. Although our limiting temperature is too large to determine the asymptotic behavior for $\lambda^{-2}(T\rightarrow0)$, it may be asserted that, for the higher dopings in Figure~\ref{lambda2}, the curvature at low temperature is too weak to allow for a fit with  a clean isotropic $s$-wave model, unlike for the lower doping (Fig.~\ref{lambda2}, bottom  inset). In the case of the strongest  doping, quasi-linear behavior does not either allow for a fit using  a $d$-wave model. Such a quasi--linear behavior may be obtained down to $T_c\approx 0.3$, as observed here, provided there is a sufficiently small gap on the Fermi surface. This is the case of the anisotropic $s$-wave model. Restricting ourselves to the weak--coupling limit
and  standard four-fold asymmetry, the gap can be expressed as $\Delta(T,\varphi)=\Delta_0\Delta(T/T_c)(1+a\cos(4\varphi))/(1+a)$, where $\varphi$ is the angle within the planes, $a\leq1$ measures the anisotropy,  $\Delta_0=1.76 k_B T_c (1+a)(1-3a/4)$ is the maximum gap on the Fermi surface\cite{Clem1966} and $\Delta(T/T_c)$ is the reduced BCS temperature dependence\cite{Carrington2003}. A reasonable fit is obtained for our sample with higher doping, using  $a = 0.65 \pm 0.05$; this large $a$ implies that the smallest value of the gap at the Fermi surface is only 20\% of $\Delta_0$. Introducing strong coupling effects would increase the value of $a$.

Ref.~\onlinecite{Das2008} proposes a competing two--band model: 
the behavior close to linear for $T/T_c > 0.25$ which has been reported for overdoped PCCO is attributed to the merging of an electron and a hole pocket of the Fermi surface into a single nodal hole pocket as the doping increases. Due to a similar evolution of the Fermi surface with doping, an analogous behavior might also be expected in the case of Nd$_{2-x}$Ce$_x$CuO$_4$  (NCCO)\cite{Das2008,Kusko2002}. In the present case, a close examination of $\lambda^{-2}(T)$ (Fig.~\ref{lambda2}, top inset) reveals further similarity with PCCO: for the higher doping also, $\lambda^{-2}(T)$ exhibits an \textit{upwards} curvature for the higher temperatures, and a \textit{downwards} one for the lower ones (disregarding the regime close to $T_c$, which may be influenced by intrinsic or extrinsic factors). \textit{A contrario}, both optimally doped PCCO and the lower SLCO doping state appear to show a \textit{downwards} curvature in the whole temperature range. Within the two--bands model, the change in curvature for $\lambda^{-2}(T)$ arises from the mixing of the contribution to the superfluid density of the nodal hole pocket (showing upwards curvature) and of the one of the anti--nodal electron pocket (showing downwards curvature)\cite{Das2008}.

\begin{figure}
\resizebox{0.9\columnwidth}{!}{\includegraphics{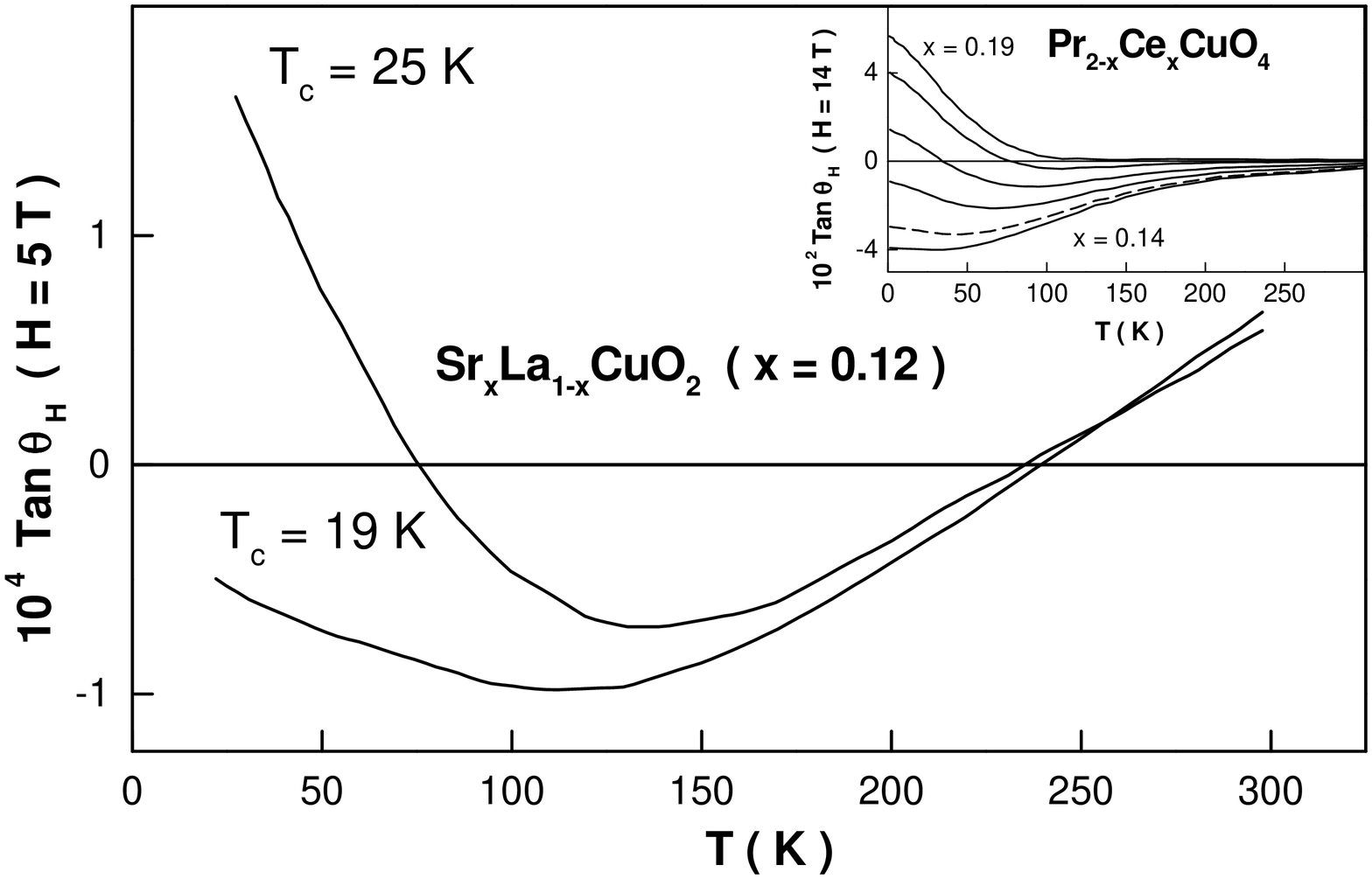}}
\caption {Hall angle for underdoped SLCO, as obtained from the data in Refs.~\onlinecite{Jovanovic2009,Jovanovic2010}. Inset : for PCCO, as computed from a mixing of underdoped and overdoped experimental data\cite{Dagan2007,Das2008} ($x = 0.14, 0.15, 0.155, 0.16, 0.18, 0.19)$; the dashed line is for optimal doping).}\label{lambda1}
\end{figure}

Additional evidence for a contribution of a hole pocket is found from the comparison of the electronic transport properties for both materials. As can be seen in Fig.~\ref{lambda1}, the Hall angle for SLCO contains a positive contribution that becomes larger as the temperature decreases and becomes positive for the highest $T_c$. A very similar behavior was observed in PCCO (see figure~\ref{lambda1}), even though the value of the Hall effect is smaller by two orders of magnitude, indicating a much lower scattering. This behavior of the Hall angle at low temperature was interpreted as the signature of the hole pocket existence at the Fermi surface. Despite the difference of value, the similarity of the temperature dependence of the Hall effect in our samples is consistent with the existence of such a hole pocket in SLCO, with a contribution that grows larger as the doping is increased. Thus, while the monotonic increase of $T_c$ with doping suggests that all our samples are still in the underdoped state, data in Fig.~\ref{lambda1} show a similar behavior to what is observed in \textit{overdoped} PCCO.

One may first question these observations as being intrinsic properties of SLCO. Our films may differ from the bulk material in several ways. First, SLCO films epitaxially grown on (100) KTaO$_3$ substrates are likely highly stressed. The substrate parameter for KTaO$_3$ is 3.989 ~\AA, while the basal plane parameter reported  for bulk SLCO is  $a = b = 3.95$~\AA\cite{Jorgensen1993}. Large parameter mismatch may have no consequence for soft materials;  for instance, Bi-based cuprates (basal parameter 3.79--3.83~\AA, Young's modulus $E \approx 40$~GPa \cite{Ye1990}) may be epitaxially grown on a variety of substrates, ranging from SrTiO$_3$, $a = 3.905$~\AA, to MgO, $a=4.21$~\AA. However,  the compact crystallographic cell of SLCO would rather indicate a large Young's modulus. Indeed, we have noticed that our films, if  submitted to a local mechanical stress, may delaminate, leaving a patchwork of free standing and epitaxial film zones, that in turn induce local twinning of the substrate. Such an observation is usually a manifestation of highly stressed films. 

The  Young's modulus of one of our films has been determined from nano-indentation measurements, using Oliver and Pharr elasto-plastic model\cite{Oliver1992}, yielding  $E=280 \pm 10$~GPa. The small thickness of the films probably does not allow the stress to relax.  Indeed, assuming a uniformly stressed, isotropic film and taking for the strain in the basal plane the difference between the substrate parameter and the bulk-SLCO  parameter, ($\epsilon_\parallel = 9.9~10^{-3}$), and for the transverse one the difference between the bulk value $c = 3.42$~\AA~ and the one measured for our films, $c=3.398$~\AA, ($\epsilon_{\bot}  = - 6.4~10^{-3}$), we obtain  the Poisson ratio $\nu = - (\epsilon_\bot / \epsilon_{\parallel})/(2-\epsilon_\bot / \epsilon_{\parallel}) = 0.24$, which is quite a reasonable value, as compared to simple oxides or cuprates\cite{Poisson}. 

\begin{figure}
\resizebox{0.9\columnwidth}{!}{\includegraphics{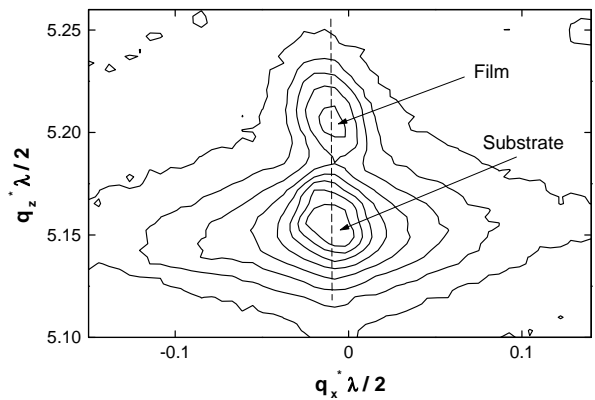}}
\caption {Contour plot (log. scale) for a $\omega$--$2\theta$ scan in the vicinity of the (411) reflection, showing alignment of the film and substrate peaks, indicative of a fully constrained film, for a 400 \AA~thick film.  Dashed line is a guide to the eye.}\label{lambda8}
\end{figure}

In addition, we performed $\omega - 2\theta$ X-ray diffraction scans. They showed an  alignment of the substrate and the film peaks, which is  characteristic of an unrelaxed epitaxial thin film (Fig.~\ref{lambda8}). This is in line with several observations showing  that oxides films need a much larger thickness to relax  than would be predicted from thermodynamic models\cite{Ingel2002}. Using the modulus measured for our films, a uniform stress about 3~GPa is expected, which may modify the band structure from relaxed bulk SLCO. The simplest effect for such a band structure modification would likely be to shift the doping state (towards higher doping, as we have seen above); however, we failed to observe a decrease of $T_c$ with doping, as would be expected in the overdoped regime.

Then, doping with oxygen may not be equivalent to doping with Sr/La substitution. Additional  oxygen atoms into the Sr$_{1-x}$La$_x$ layers  also introduce Cu-O bonds between CuO$_2$ planes, via the apical oxygen, that are not present in the original material. While it is known that oxygen vacancies present in distant charge-reservoir planes or chains may locally alter the electronic properties of the conducting planes\cite{McElroy2005}, such a proximity of the doping atom may have dramatic effects, eventually resulting in inhomogeneous superconductivity. In the case of Sr$_2$CuO$_{3+x}$, it was shown that ordering of the apical oxygen has a noticeable effect on the superconducting transition temperature, in the absence of a change of the doping in the CuO$_2$ planes\cite{Liu2006}, as also has disorder in the adjacent SrO plane for Bi-based compounds\cite{Fujita2005}. Previous studies have shown that our SLCO films with  larger oxygen content have larger transition width, as the zero resistance temperature experiences larger shift with oxygen content than the onset temperature does. Within this perspective, it is reasonable to assume that  oxygen content  alters  both doping and disorder.  Disorder can have a strong effect on both the superconducting temperature and the superfluid density, when nodes are present in the order parameter. It has indeed be noticed, in the case of PCCO, that the values computed for $\lambda(0)$ are well below the experimental ones, suggesting  also a  doping-induced disorder \cite{Das2007}. For SLCO, the effect should be stronger, due to the specific position of the doping oxygen.

\begin{table}
\caption{\label{tab:table1} $\left(\text{d}\lambda^{-2}(T)/\text{d}T\right)_{T_c}$ and extrapolated values for $\lambda(T=0)$, as explained in the text.}
\begin{ruledtabular}
\begin{tabular}{lcr}
$T_c$&$-10^2\,\left(d\lambda^{-2}(T)/dT\right)_{T_c}$ &$\lambda(T=0)$ \\
(K) & ($\mu$m$^{-2}$K$^{-1}$) & ($\mu$m)\\
\hline
7.1 & 0.43 & 7.1\\
10.8 & 2.4 & 2.2\\
12 & 2.6 & 2.1\\
13 & 6.8 & 1.2\\
14.3 & 8.1 & 1.0\\
14.9 & 11.3 & 0.85\\
19 & 11.4 & 0.69\\
\end{tabular}
\end{ruledtabular}
\end{table}

Given that our samples are strongly disordered, the behavior of $\lambda(T)^{-2}$ should be affected.  It is indeed well known that, for a $d$-wave superconductor, scattering induces a finite density of states at the Fermi level and changes the zero-temperature asymptotic behavior from a linear to a quadratic law in temperature,  whereas, above some crossover temperature, the pure regime is recovered\cite{Hirschfeld1993}. By analogy, within the two--bands model, one would expect the contribution of the hole pocket that carries the d--wave character of the superconductivity to be strongly affected by disorder. However, our measurements do not cover the low-temperature range where disorder should dominate the $\lambda^{-2}$ behavior. Indeed, out analysis is made for $T/T_c \geq 0.3$,  where the contribution of disorder is expected to be negligible. 
There are indications, from the available data on cuprates with a simpler Fermi surface, that one may simultaneously observe a strong reduction of both $T_c$ and the superfluid density due to disorder, and a temperature behavior for $\lambda(T)$ reminiscent of their $d$-wave character, in such a high-temperature regime (see e.g. Ref.~\onlinecite{Ulm1995}, where a YBa$_2$CuO$_7$ thin film substituted by 6\% Ni exhibits superfuid density reduced by a factor 25 and $T_c$ reduced by a factor 1.3, while above $T/T_c \simeq 0.3$, the pure $d$-wave result provides a good fit to the data; see also  Ref.~\onlinecite{Salluzzo2001}). Although the Fermi surface of our electron-doped compound is  more complex than that of hole-doped cuprates, we similarly expect that the high-temperature pure behavior is preserved in the present case. Thus, although the details of the Fermi surface also contribute to the high temperature behavior of the superfluid density and may introduce some discrepancy with respect to the conventional $d$-wave result for a single cylindrical Fermi surface, we consider the strong departure from the $s$-wave result (Fig.~\ref{lambda2}) as a possible contribution from a nodal band.

\begin{figure}
\resizebox{0.9\columnwidth}{!}{\includegraphics{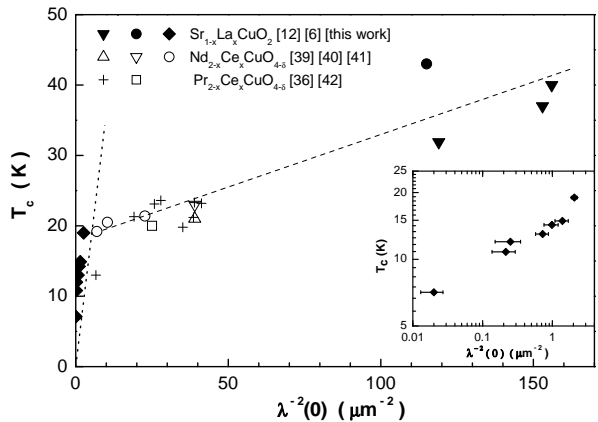}}
\caption {$T_c$ vs $\lambda^{-2}(T=0)$ for various e-doped cuprates. The dotted line is the Uemura line for moderately underdoped hole-doped cuprates\cite{Uemura1989}; the dashed line is a guide to the eye. Inset: log-log representation of our data, where the error bars are estimated from the different methods of $\lambda(0)$ extrapolation. These data follow $T_c \propto (\lambda^2(0))^{-0.2}$.}\label{lambda6}
\end{figure}

\begin{figure}
\resizebox{0.9\columnwidth}{!}{\includegraphics{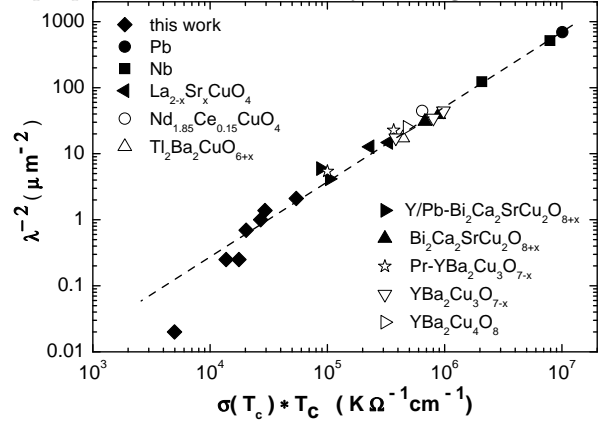}}
\caption {Full circles : universal scaling for various cuprates and metals, as proposed in Ref.~\onlinecite{Homes2004} (in--plane conductivities, from the supplementary information). Diamonds : this work (the conductivity is estimated from the data in Ref.~\onlinecite{Jovanovic2009}). Dashed line is a guide to eye.}\label{lambda10}
\end{figure}

Finally, we comment on the $T_c$--$\lambda^{-2}(0)$ relationship. Even though the lowest available temperature is 4~K, the temperature range of our experiment is enough to allow the estimation of $\lambda^{-2}(0)$ by extrapolating $\lambda^{-2}(T)$. This may done either using a linear extrapolation or a quadratic one, which yields close results (the averaged values are presented in Table~\ref{tab:table1} and Fig.~\ref{lambda6}). These extrapolated values for $\lambda(0)$ (Fig.~\ref{lambda6}) are much larger than what has been previously measured in bulk SLCO near optimal doping\cite{Khasanov2008,Satoh2008}. $T_c$ is found to extrapolate to zero as $(\lambda^{-2}(0))^\alpha$, where $\alpha \approx$~0.2 (Fig.~\ref{lambda6}, inset).
The data strongly suggest a crossover to a phase-fluctuation regime, as $1/\lambda(0)^2 \rightarrow 0$, when the superfluid density is low enough to impose a superconducting transition driven by phase ordering. The fluctuations may be of thermal origin or driven by the proximity of a quantum critical point (QCP)\cite{Emery1995}. In the case of a QCP, a sub-linear dependence is expected, as observed in our case (Fig.~\ref{lambda6}, inset). However, both quantities are expected to be related as $T_c \propto (1/\lambda(0)^2)^\alpha$, where $\alpha=z/(z+D-2)$, $z \geq 1$ (see e.g. Ref.~\onlinecite{Hetel2007} and Refs therein) and $D=3$ in the present case, owing to comparable interplane distance and coherence length\cite{Jovanovic2009}. The resulting value, $z \approx 0.25$, is an unphysically small, making the QCP scenario unlikely. The phase-fluctuation scenario should be favored by weak phase stiffness and a short coherence length. SLCO (as well as NCCO\cite{Kim2003}) shows a smaller upper critical field than hole-doped cuprates ($\text{d}B_{c2}/\text{d}T \approx$ 0.3 T~K$^{-1}$ \cite{Khasanov2008}; 0.5 T~K$^{-1}$  \cite{Jovanovic2009}, while $\text{d}B_{c2}/\text{d}T \approx$ 2 T~K$^{-1}$ for hole-doped cuprates) and a relatively small penetration depth (Fig.~\ref{lambda6}) : this does not make it a likely candidate for the phase-fluctuation scenario, as found in Ref.~\onlinecite{Satoh2008}. However, with decreasing doping in the e-doped materials, the antinodal carriers become dominant, for which there is a finite coherence length (as opposed to the nodal direction), while there is an increase of the superfluid density (due to the reduction in the carrier density and/or stronger disorder): both effects could then favor a crossover from a conventional mean-field behavior to a superconducting transition driven by phase fluctuations. The linear $T_c$--$\lambda^{-2}(0)$ relationship, which is thought to be characteristic of this mechanism\cite{Uemura1989}, is however also not observed by us. Finally, an alternative universal scaling was proposed in Ref.~\onlinecite{Homes2004}, linearly relating the superfluid density to the product $\sigma_{DC} T_c$. Alternatively, it may be viewed as relating the scattering rate at $T_c$ to this parameter. We observe that our data -- with the exception of the sample with the lower $T_c$ -- obey such a scaling reasonably well (Fig.~\ref{lambda10}), being situated at the opposite of the large $\sigma_{DC}\,T_c$ product of metal superconductors.

In summary, the penetration depth of thin films infinite-layer Sr$_{0.88}$La$_{0.12}$CuO$_{2+x}$ thin films is found to depart from the isotropic $s$-wave behavior as doping is increased, indicating the contribution of a smaller gap on the Fermi surface. Both an anisotropic $s$-wave order parameter and a two-band model, as was used for overdoped Pr$_{2-x}$Ce$_x$CuO$_4$, may account for the data. In the latter case, the small gap originates from a nodal hole pocket with a $d$-wave character\cite{Das2008}. Large values of the zero-temperature penetration depth are observed. Disorder on the apical oxygen site and the associated strong scattering in the CuO$_2$ plane may be the primary cause for the superfluid density reduction.

\begin{acknowledgments}
The authors are grateful to T.R. Lemberger for sharing valuable knowledge on the ac susceptibility technique. 
The Orsay group acknowledge the support of the A.N.R. under project No.ANR-07--1--19--3024.
\end{acknowledgments}

\newpage


\end{document}